\documentclass[epj]{svjour}
\newcommand{\app}{\setcounter{section}{1}
\setcounter{equation}{0} \renewcommand{\thesection}{APPENDIX
\Alph{section}}\renewcommand{\theequation}{\Alph{section}.\arabic{equation}}}
\usepackage{latexsym}
\usepackage{amsmath}
\usepackage{graphics}
\newcommand{\qln}{ \ln_{q} }
\newcommand{\qexp}{ \exp_{q} }

\begin{document}
\title{A non self-referential expression of Tsallis' probability
distribution function}
\author{T. Wada\inst{1} \and A.M. Scarfone\inst{2}% etc
}                     % Do not remove
\institute{Department of Electrical and Electronic Engineering,
Ibaraki University, Hitachi,~Ibaraki, 316-8511, Japan.\\
\email{wada@ee.ibaraki.ac.jp} \and Istituto Nazionale di Fisica
della Materia (INFM) and Dipartimento di Fisica,
 Politecnico di Torino,\\ Corso Duca degli Abruzzi 24, 10129
Torino, Italy.\\
\email{antonio.scarfone@polito.it}}

\date{Received: date / Revised version: date}
\abstract{ The canonical probability distribution function (pdf)
obtained by optimizing the Tsallis entropy under the linear mean
energy constraint (first formalism) or the escort mean energy
constraint (third formalism) suffer self-referentiality. In a
recent paper [Phys. Lett. A {\bf335} (2005) 351-362] the authors
have shown that the pdfs obtained in the two formalisms are
equivalent to the pdf in non self-referential form. Based on this
result we derive an alternative expression, which is non
self-referential, for the Tsallis distributions in both first and
third formalisms.
\PACS{
      {05.20.-y}{Classical statistical mechanics}   \and
      {05.90.+m}{Other topics in statistical physics}
     } % end of PACS codes
} %end of abstract
\maketitle
\section{Introduction}
\label{intro}

Since the pioneering work of Tsallis \cite{Tsallis88}, a large
number of papers have been published based on the framework of the
so-called nonextensive thermostatistics
\cite{Tsallis98,NEXT01,NEXT03}.\\ Tsallis' entropy is a
generalization of the Boltzmann-Gibbs (BG) entropy defined by
\begin{equation}
 S_q \equiv \sum_i \left( \frac{p_i^q-p_i}{1-q} \right),
\label{ent}
\end{equation}
where $p_i$ is the probability of $i$-th state of the system and
$0\leq q\leq2$ is a real deformed parameter. Here, for the sake of
simplicity, the Boltzmann constant is set to unity. In the $q \to
1$ limit, $S_q$ reduces to the standard BG-entropy $S = -\sum_i
p_i \ln (p_i)$.

The current formulation of Tsallis thermostatistics has been
established in Ref. \cite{Tsallis98}. In the first formalism the
probability distribution function (pdf) is obtained by maximizing
$S_q$ with the conditions posed by the normalization
\begin{equation}
\sum_i p_i=1 \ ,\label{norm}
\end{equation}
and by the linear average energy
\begin{equation}
\sum_i p_i\,E_i=U \ ,\label{lenergy}
\end{equation}
as follows
\begin{equation}
 \frac{\delta}{\delta p_i}
 \left(
    S_{q} - \bar\beta\,\sum_j p_j\,E_j
   -\bar\gamma \sum_j p_j
 \right) = 0 \ ,
  \label{MaxEnt1}
\end{equation}
where $\bar\gamma$ and $\bar\beta$ are the Lagrange multipliers
associated to the constraints (\ref{norm}) and (\ref{lenergy}),
respectively.\\
The solution of this MaxEnt problem is
\begin{align}
  p_i = \frac{1}{\bar{Z}_q^{\rm (1)}} \cdot
   \exp_{2-q}\left( \frac{-\bar{\beta} (E_i - U)}
      {q\,\sum_i p_i^q} \right) \ ,
  \label{pdf1}
\end{align}
where $\bar{Z}_q^{\rm (1)}$ is the generalized partition function
in the first formalism, and the following relation is satisfied.
\begin{equation}
  \bar{Z}_q^{(1)}=\left(\sum_ip_i^q\right)^{\frac{1}{1-q}} \ .\label{z1}
\end{equation}
$\bar{Z}_q^{\rm (1)}$ is also related to the Lagrange multiplier
$\bar\gamma$ through the relation
\begin{equation}
1+\bar\gamma=q\,\ln_q \bar Z_q^{\rm (1)} -\bar\beta\,U \
.\label{Z}
\end{equation}
In Eqs. (\ref{pdf1}) and (\ref{Z}) we have introduced the
$q$-deformed exponential
\begin{equation}
\qexp(x)=[1+(1-q)\,x]^{1\over1-q} \ ,
\end{equation}
and its inverse function the $q$-deformed logarithm
\begin{equation}
\qln(x)={x^{1-q}-1\over1-q} \ .
\end{equation}
Note that Eq. \eqref{pdf1} is the pdf which Di Sisto {\it et al.}
\cite{Sisto} and Bashkirov \cite{Bashkirov} have independently
obtained by modifying the treatment of the original version of
Tsallis \cite{Tsallis88}.

On the other hand, in the third formalism the Tsallis pdf is
obtained by maximizing $S_q$ with the conditions posed by the
normalization (\ref{norm}) and by the normalized $q$-average
energy
\begin{equation}
\frac{\sum_i p_i^q\,E_i}{\sum_i p_i^q}= U_q \ ,\label{energy}
\end{equation}
as follows
\begin{equation}
 \frac{\delta}{\delta p_i}
 \left(
    S_{q} - \beta\,\frac{\sum_j p_j^q\,E_j}{\sum_k p_k^q}
   -\gamma\,\sum_j p_j
 \right) = 0 \ ,
  \label{MaxEnt3}
\end{equation}
where $\gamma$ and $\beta$ are the Lagrange multipliers associated
to the constraints (\ref{norm}) and (\ref{energy}),
respectively.\\
>From Eq. (\ref{MaxEnt3}) we obtain
\begin{align}
  p_i = \frac{1}{\bar{Z}_q^{(3)}}
   \cdot\qexp \left( -\frac{\beta\, (E_i - U_q)}{\sum_j p_j^q}
 \right) \ ,
 \label{p3}
\end{align}
where $\bar{Z}_q^{(3)}$ is the generalized partition function in
the third formalism, and
\begin{equation}
  \bar{Z}_q^{(3)}=\left(\sum_ip_i^q\right)^{1\over 1-q}.
  \label{Zq}
\end{equation}
Both expressions (\ref{pdf1}) and (\ref{p3}) are explicitly
self-referential, which cause some difficulties. One of them is
concerning with numerical convergence when one calculates the
$p_i$. In order to overcome this problem, Tsallis {\it et al.}
\cite{Tsallis98} proposed the two different calculation methods:
``the iterative procedure''; and ``$\beta \to \beta'$
transformation''. Both methods were further studied by Lima and
Penna \cite{Lima-Penna99}.\\
An even more serious difficulty is that a maximum is not
necessarily guaranteed due to the nondiagonal Hessian matrix
associated to the above Lagrange procedure. In order to overcome
this difficulty Mart\'\i nez {\it et al.} \cite{Martinez00} have
introduced an alternative Lagrange route, which is the so called
optimal Lagrange multiplier (OLM) formalism
\cite{0th-law,Abe-Martinez01,Abe01,Casas02}.

In our previous work \cite{Wada-Scarfone05}, the equivalences of
some different expressions of the Tsallis pdfs have been studied.
Among them, it was explicitly shown that the non self-referential
pdf
\begin{equation}
  p_i = \alpha\,
        \qexp \left( -\bar{\beta}\,E_i - \bar{\gamma} \right) \ ,
 \label{pdf}
\end{equation}
with $\alpha=(2-q)^{1/(q-1)}$, maximizes the entropy
\begin{equation}
S_{2-q}=-\sum_ip_i\qln(p_i)\equiv\sum_i\left({p_i^{2-q}-p_i\over
q-1}\right) \ ,
\end{equation}
under the constraints (\ref{norm}) and (\ref{lenergy}).
Furthermore it was shown that Eq. \eqref{pdf} is equivalent to the
self-referential forms (\ref{pdf1}) and (\ref{p3}) arising in the
first and third formalism, respectively.\\ On the base of these
equivalences, we can obtain a non self-referential expression of
the Tsallis pdf not only for the first formalism but also for the
third formalism. This is the purpose of the present note.

\section{The non self-referential expressions}
\label{sec:non self-ref}

We begin with the first formalism. Let us briefly review the
equivalence between the pdf in the form (\ref{pdf}) and the
distribution (\ref{pdf1}). From the MaxEnt problem
\begin{equation}
 \frac{\delta}{\delta p_i}
 \left(
    S_{2-q} - \bar\beta\,\sum_j p_j\,E_j
   -\bar\gamma \sum_j p_j
 \right) = 0 \ ,
  \label{MaxEnt21}
\end{equation}
we obtain
\begin{equation}
  \frac{(2-q)p_i^{1-q}-1}{q-1} -\bar\beta\,E_i-\bar\gamma=0\ . \label{s1}
\end{equation}
Multiplying the both sides by $p_i$ and summing over $i$, taking
into account of the constraints (\ref{norm}) and (\ref{lenergy}),
it follows the solution \cite{Wada-Scarfone05}
\begin{align}
  p_i = \frac{1}{\bar{Z}_{2-q}^{\rm (1)}} \cdot
   \exp_q\left( \frac{-\bar{\beta} (E_i - U)}
      {(2-q)\,\sum_i p_i^{2-q}} \right) \ ,
  \label{eq1}
\end{align}
where
\begin{equation}
{1\over\bar{Z}_{2-q}^{(1)}}=\alpha\,\qexp\left(-\bar\gamma-\bar\beta\,U\right)
\ ,
\end{equation}
is equivalent to Eq. (\ref{Z}) with $q$ replaced by $2-q$. Note
that by replacing $q$ with $2-q$ in Eq. (\ref{eq1}), we recover
the form of Eq. (\ref{pdf1}).\\
On the other hand, Eq. (\ref{s1}) can be immediately solved w.r.t.
$p_i$ as
\begin{equation}
   p_i=\qexp\left(-\frac{1+\bar\gamma+\bar\beta\,E_i}{2-q}\right) \,
\label{ns1}
\end{equation}
%and according to Eq. (\ref{Z}), Eq. (\ref{ns1}) can be written in
%\begin{equation}
%p_i=\qexp\left(-\frac{q}{2-q}\,\qln \bar
%Z_q^{(1)}-\frac{\bar\beta}{2-q}\,(E_i-U)\right) \ ,
%\end{equation}
which is equivalent to the non self-referential expression
\eqref{pdf} in the first formalism.

We consider now the expression of the pdf in the third formalism.
%The equivalence with Eq. (\ref{pdf}) is readily
%obtained by multiplying Eq. (\ref{s1}) by $p_i^q/\sum_jp_j^q$ and
%summing over $i$. Taking into account of the constraints
%(\ref{norm}) and (\ref{energy}), and by employing the following
%relation \cite{Wada-Scarfone05}
%\begin{equation}
%\beta={\left(\sum_ip_i^q\right)^2\over2-q}\,\bar\beta \ ,
%\end{equation}
%we obtain the pdf of the third version in the form given by Eq.
%(\ref{p3}). In this case it can be shown that the canonical
%partition function $\bar{Z}_q^{(3)}$ is related to the Lagrange
%multipliers $\bar\gamma$ and $\bar\beta$ trough the relation
%\begin{equation}
%{1\over\bar{Z}_q^{(3)}}=\alpha\,\qexp\left(-\bar\gamma-\bar\beta\,U_q\right)
%\ ,
%\end{equation}
%which is equivalent to the definition (\ref{Z3}).\\
The derivation of the non self-referential expression equivalent
to the pdf (\ref{p3}) is very simple, and the point is to express
the quantity $\sum_j p_j^q$ in terms of $\gamma$. From Eq.
\eqref{MaxEnt3} we have
\begin{equation}
 \frac{q p_i^{q-1}-1}{1-q}
 - \beta \frac{ q p_i^{q-1}}{\sum_j p_j^q}
        \left(E_i - U_q \right)
   -\gamma = 0 \ .
  \label{MaxEnt3-rel}
\end{equation}
Multiplying the both sides of this equation by $p_i$ and taking
summation over $i$, we obtain
\begin{equation}
   \sum_i p_i^q
     = \frac{1+(1-q) \gamma}{q}
     = 1 + (1-q)\left( \frac{\gamma + 1}{q} \right) \ .
\end{equation}
From Eq. \eqref{Zq} this relation is also expressed as
\begin{equation}
  \bar{Z}_q^{(3)} = \qexp \left( \frac{\gamma+1}{q} \right).
  \label{Zq-gamma}
\end{equation}
By utilizing Eq. \eqref{Zq}, Eq. \eqref{p3} can be written as
\begin{align}
  p_i &= \frac{1}{\bar{Z}_q^{(3)}}
   \; \qexp \left(-\frac{\beta}{\left(\bar{Z}_q^{(3)}\right)^{1-q}}\,
   (E_i - U_q) \right)
\nonumber \\
 &= \frac{1}{\left( \bar{Z}_q^{(3)} \right)^2 }
   \; \qexp \left( \qln \bar{Z}_q^{(3)} -\beta (E_i - U_q)\;
   \right) \ ,
\end{align}
Substituting Eq. \eqref{Zq-gamma} into this we finally obtain
\begin{align}
  p_i &= \left[ \qexp\left(\frac{\gamma+1}{q}\right) \right]^{-2}
   \; \qexp \left( \frac{\gamma+1}{q} -\beta (E_i - U_q)\;
   \right).
  \label{NSR}
\end{align}
We remark that $\gamma$ depends on only $U_q$ and $\beta$. Eq.
\eqref{NSR} is thus the non self-referential expression of the pdf
in the framework of the third formulation.

\section{Conclusions}
\label{conclusions}

We have obtained the non self-referential expressions of the
Tsallis pdf for both first and third formalisms based on the
equivalences among the expression given in Eq. (\ref{pdf}), which
is non self-referential, the ones given in Eqs. (\ref{pdf1}) and
(\ref{p3}), which are self-referential. These non self-referential
expressions permit us to overcome some computational problems as
well as to simplify the treatment of other related difficulties
which arise from employing the self-referenced expressions
(\ref{pdf1}) and (\ref{p3}). Notwithstanding, we remark that
although implicitly defined, both pdfs (\ref{pdf1}) and (\ref{p3})
remain still useful in the analytical computations of many
properties.

\section*{Note added}

We would like to thank Dr. Suyari for informing his work
concerning on this issue. During proceeding this work, T.W. had a
chance to know that Dr. Suyari has independently performed the
similar work \cite{Suyari05}. After almost completing this work,
Dr. Suyari and we exchanged the results each other, and noticed
that both he and we obtained the same results by different
methods.

\section*{Appendix}
\app
Let us show here the equivalence between Eq. \eqref{NSR} and Eq.
(30) in Ref. \cite{Suyari05}, which can be written in our notation
as follows.
\begin{align}
   p_i = \frac{1}{\bar{Z}_q^{(3)}}
   \; \qexp \Big( -\beta_q (E_i - U_q) \Big) \ ,
  \label{Suyari's pdf}
\end{align}
where
\begin{equation}
  \beta_q \equiv \frac{\beta}{1+(1-q)\left( \frac{\gamma+1}{q} \right)} \ .
\end{equation}
By utilizing the identity
\begin{equation}
  \qexp(x+y) = \qexp(x) \qexp\left( \frac{y}{1+(1-q)x} \right) \ ,
\end{equation}
it follows
\begin{align}
&\qexp \left( \frac{\gamma+1}{q} -\beta (E_i - U_q)\;\right)
  \qquad \qquad \qquad \nonumber \\
  =&\qexp\left(\frac{\gamma+1}{q}\right) \cdot
  \qexp \left( -\beta_q (E_i - U_q)\;\right) \ .
\end{align}
Substituting this  into Eq. (\ref{NSR}) we obtain
\begin{align}
p_i =\left[ \qexp\left(\frac{\gamma+1}{q}\right) \right]^{-1}
   \cdot
  \qexp \left( -\beta_q (E_i - U_q)\;\right) \ ,
\end{align}
which is equivalent to \eqref{Suyari's pdf} as can be see by
taking into account of Eq. \eqref{Zq-gamma}.

%\begin{figure}
%\resizebox{0.75\textwidth}{!}{%
%  \includegraphics{leer.eps}
%\vspace{5cm}       % Give the correct figure height in cm
%\end{figure}
%\begin{figure*}
%\vspace*{5cm}       % Give the correct figure height in cm
%\end{figure*}
%\begin{table}
%\begin{tabular}{lll}
%\hline\noalign{\smallskip}
%\end{tabular}
% Or use
%\vspace*{5cm}  % with the correct table height
%\end{table}
%
% BibTeX users please use
% \bibliographystyle{}
% \bibliography{}

\end{document}